\documentclass[twocolumn,showpacs,preprintnumbers]{revtex4}
\usepackage{graphicx}
\usepackage{dcolumn}
\usepackage{bm}

\begin{document}

\title{Decoherence in a superconducting flux qubit with a $\pi$-junction}
\author{T. Kato$^1$, A. A. Golubov$^2$, and Y. Nakamura$^{3,4}$
\footnote{Submitted to Phys. Rev. B}}
\affiliation{$^1$ Institute for Solid State Physics, The University of Tokyo, Kashiwa,
Chiba 277-8581, Japan \\
$^2$ Faculty of Science and Technology, University of Twente, 7500 AE
Enschede, The Netherlands\\
$^3$ Nano Electronics Research Laboratories, NEC Corporation, Tsukuba, Ibaraki 
305-8501, Japan\\
$^4$ Frontier Research System, The Institute of Physical and Chemical Research (RIKEN), 
Wako, Saitama 351-0198, Japan
}

\begin{abstract}
We consider the use of a $\pi$-junction for flux qubits to realize
degenerate quantum levels without external magnetic field. On the
basis of the Caldeira-Leggett model, we derive an effective
spin-Boson model, and study decoherece of this type of qubits. We
estimate the dephasing time by using parameters from recent
experiments of SIFS junctions, and show
that high critical current and large subgap resistance are
required for the $\pi$-junction to realize a long coherent time.
\end{abstract}

\date{\today }
\pacs{03.67.Lx, 03.65.Yz, 85.25.Cp, 85.25.Dq}
\maketitle




It is now well established that in addition to conventional Josephson
junctions having an energy minimum at zero phase difference across the
junction, there exist the so-called $\pi$-junctions which provide the phase
shift of $\pi $ in the ground state. The intrinsic $\pi$-shifts were first
realized in grain boundary Josephson junctions in $d$-wave 
superconductors~\cite{Tsuei94,Tsuei00}. Subsequently, $\pi$-junctions have been
realized in hybrid structures between high-$T_{\mathrm{c}}$ and
low-$T_{\mathrm{c}}$ superconductors~\cite{Tsuei00,high_Tc_low_Tc}
and by injection of quasi-particles~\cite{quasi_particle_injection}.
Recent development in fabrication of
superconductor-ferromagnet-superconductor (SFS) junction made
it possible to obtain a $\pi$-junction with high critical current density~\cite{SFS}.
An advantage of SFS junctions is the possibility to combine them with usual
low-$T_{\mathrm{c}}$ superconductive circuits using conventional fabrication
technique.

The use of $\pi$-junctions provides several new applications. For example,
the application of $\pi $-junctions as complementary devices in SFQ logic
was recently proposed~\cite{logic} and realized in high-$T_{%
\mathrm{c}}$-low-$T_{\mathrm{c}}$ junctions~\cite{Ortlepp06}. It is
interesting that before this `classical' application of the $\pi$-junction,
the use of $\pi$-junctions for realization of quantum two-state systems was
considered~\cite{quiet_qubit}. In this qubit system, the $\pi$%
-junction was used as a $\pi$ phase shifter along the loop instead
of current biasing or external magnetic flux. After this proposal,
remarkable progress in fabrication, coherent control of one qubit,
controllable coupling between qubits, and readout with high fidelity
has been achieved in superconducting qubits~\cite {charge_qubit,
quantronium_qubit, flux_qubit, Chiorescu03,phase_qubit}.
Nevertheless, up to now, the use of $\pi$-junctions to qubits has
not been studied experimentally. One of difficulties for realization
may lie on the original proposal in which a qubit consists of
complicated circuits with many Josephson
junctions~\cite{quiet_qubit}. Another serious difficulty comes from
dissipation due to quasi-particle excitation, which is unavoidable
in many realizations of $\pi$-junctions. Generally, qubits suffer
strong decoherence by excitation in the environment. 

\begin{figure}[tb]
\begin{center}
\includegraphics[width=60mm]{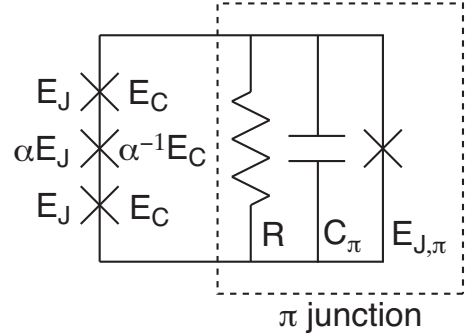}
\end{center}
\caption{Flux qubit circuits with a $\pi$-junction. Shunt resistance at the
$\pi$-junction is introduced for estimate of damping effects.}
\label{fig:model}
\end{figure}

In this paper, we consider the use of $\pi$-junction for phase bias of flux qubits.
The circuit we study is shown in Fig.~\ref{fig:model}. In this circuit, we need
no external flux to realize degenerate quantum levels, because the phase drop across the
three Josephson junction is adjusted as $\pi$ by the $\pi$-junction with a
large Josephson energy. This type of phase bias can avoid dephasing due to
noise in external flux, and is frequently called as a `quiet qubit'.
In actual experiments, however,  damping at the $\pi$-junction may cause
severe decoherence on the qubit. The purpose of this paper is to derive the effective
spin-Boson model describing the flux qubit with a damped $\pi$-junction, and
to estimate the dephasing time by using realistic experimental parameters.
We clarify the condition for long coherence time in this qubit system, and discuss
the possibility of the use of $\pi$-junctions for qubits by referring recent
experiments on SIFS junctions.

In order to describe damped dynamics, we introduce the RSJ model for the
$\pi$-junction as shown in Fig.~\ref{fig:model}, where dissipation is
expressed by a resistance $R$ shunted in parallel to the junction. We expect
that this phenomenological model may give a qualitative estimate of
decoherence effects by $\pi$-junctions.
We introduce the charging energy $E_{C, \pi} = e^2/(2C_{\pi})$
and damping frequency $\gamma = 1/(RC_{\pi})$ of the $\pi$-junction.

The Hamiltonian consists of three parts as $H = H_{\mathrm{qubit}} + H_L +
H_{\pi}$. The first part $H_{\mathrm{qubit}}$ describes a flux qubit, and is
given as~\cite{Chiorescu03}
\begin{eqnarray}
H_{\mathrm{qubit}} &=& - E_{\mathrm{J}} ( \cos \phi_1 + \cos \phi_2 + \alpha
\cos \phi_3)  \nonumber \\
&+& 4 E_{\mathrm{C}} ( n_1^2 + n_2^2 + \alpha^{-1} n_3^2 ),
\end{eqnarray}
where $E_J$ is a Josephson energy, and $E_C$ is a charging energy.
Here, $\phi_i$ and $n_i$ are a phase difference and induced charge at the 
$i$-th junction, respectively. The area of one junction is reduced by the
factor $\alpha$, which is typically taken as 0.8~\cite{Chiorescu03}. The
second part of the Hamiltonian describes the inductance energy of the loop, and is given as
\begin{equation}
H_L = \frac{1}{2L} \left( \frac{\Phi_0}{2\pi} \right)^2 ( \phi_1 + \phi_2 +
\phi_3 + \phi_{\pi} - \phi_{\mathrm{ext}} )^2,
\end{equation}
where $\Phi_0 = h/(2e)$. Here, $\phi_{\pi}$ is a phase of the $\pi$
junction, and $\phi_{\mathrm{ext}} = 2\pi (\Phi_{\mathrm{ext}}/\Phi_0 )$ is
a phase induced by the external flux through the loop. By assuming small
inductance $L$, the inductive part of the Hamiltonian can be treated as a
constraint condition
\begin{equation}
\phi_1 + \phi_2 + \phi_3 + \phi_{\pi} = \phi_{\mathrm{ext}}.
\end{equation}
The third term describes the damped $\pi$-junctions, and is expressed by the
Caldeira-Leggett Hamiltonian
\begin{eqnarray}
H_{\pi} &=& + E_{J,\pi} \cos \phi_{\pi} + 4 E_{C,\pi} n_{\pi}^2  \nonumber \\
&+& \sum_{\alpha} \left\{ \frac{p_{\alpha}^2}{2m_{\alpha}} + \frac12
m_{\alpha} \omega_{\alpha}^2 \left( x_{\alpha} - \frac{C_{\alpha}}{
m_{\alpha} \omega_{\alpha}^2} \phi_{\pi} \right)^2 \right\}.
\end{eqnarray}
The damping property is determined by the spectral function
\begin{equation}
J(\omega) = \frac{\pi}{2} \sum_{\alpha} \frac{C_{\alpha}^2}{m_{\alpha}
\omega_{\alpha}} \delta(\omega - \omega_{\alpha}).
\end{equation}
In the RSJ model, the spectral function is given as
\begin{equation}
J(\omega)/M = \gamma \omega e^{-\omega/\omega_{\mathrm{c}}},
\end{equation}
where $M = 1/(8E_{C, \pi})$ is a mass of the $\pi$-junction, and $\omega_{%
\mathrm{c}}$ is a high-frequency cutoff.

In this paper, we focus on the `passive' use of the $\pi$-junction.
For this use, the Josephson energy of the $\pi$-junction should be
taken as sufficiently large. Hence, we assume $E_{J, \pi} \gg E_C,
E_J$, and approximate the Josephson energy of the $\pi$-junction as
$E_{J, \pi} (\phi_{\pi} - \pi)^2/2$. Within this approximation, the
phase of the $\pi$-junction is kept almost $\pi$. The remaining
dynamics around the potential minima is described by a damped
oscillator with a eigenfrequency $\omega_0 =
(8 E_{J, \pi} E_{C, \pi})^{1/2}/\hbar$. In the following discussion, we set $%
\hbar = 1$.

Under the condition $E_J \gg E_C$, which is taken for usual flux qubits, we
can truncate the Hamiltonian $H_{\mathrm{qubit}}$ into the two-level
Hamiltonian as $H_{\mathrm{qubit}} = H_{\mathrm{two-state}} + H_{\mathrm{%
coupling}}$. The first part $H_{\mathrm{two-state}} = (\Delta/2) \sigma_x +
(\varepsilon/2) \sigma_z$ describes the qubit system, where $\Delta$ is a
tunneling splitting, and $\varepsilon$ is a bias proportional to the
external flux $\Phi_{\mathrm{ext}}$. The second part, which describes the
coupling between the qubit and the $\pi$-junction, is given as
\begin{equation}
H_{\mathrm{coupling}} = - E_{J, \mathrm{eff}}(\alpha) \Delta \phi_{\pi}
\sigma_z,  \label{eq:coupling}
\end{equation}
where $E_{J, \mathrm{eff}} = (1 - 1/(4\alpha^2))^{1/2} E_J$, and $\Delta
\phi_{\pi} = \phi_{\pi} - \pi$.

To simplify the Hamiltonian $H_{\pi}$, we change the variables as $x =
M^{1/2} \Delta \phi_{\pi}$, $p = M^{-1/2} n_{\pi}$. We further replace the
sum in the Hamiltonian of the harmonic oscillators by the integral. This can
be performed by replacing the variables as $X_{\omega} = m^{1/2}
x_{\alpha}/(\Delta \omega)^{1/2}$, $P_{\omega} = m^{-1/2} p_{\alpha}/(\Delta
\omega)^{1/2}$, and $C_{\omega} = (M m_{\alpha})^{-1/2} C_{\alpha}/(\Delta
\omega)^{1/2}$, where $\Delta \omega$ is a length of one slice in the $%
\omega $-direction. In the limit $\Delta \omega \rightarrow 0$, we obtain
\begin{eqnarray}
& & H_{\pi} = \frac{\hat{p}^2}{2} + \frac12 \omega_0^2 \hat{x}^2  \nonumber
\\
& & + \int_0^{\infty} d\omega \left\{ \frac{P_{\omega}^2}{2} + \frac12
\omega^2 X_{\omega}^2 - C_{\omega} X_{\omega} \hat{x} + \frac{C_{\omega}}{2
\omega^2} \hat{x}^2 \right\}.
\end{eqnarray}
The coefficient $C_{\omega}$ can be related to the spectral function as
\begin{equation}
J(\omega)/M = \frac{\pi}{2} \frac{C_{\omega}^2}{\omega}.
\end{equation}

The Hamiltonian of the $\pi$-junction describing a damped oscillator can be
diagonalized exactly~\cite{Fano61,Costa00}. In order to express the
eigenmodes with the energy $\omega$, we introduce a canonical transformation
for the operators as
\begin{equation}
\bar{X}_{\omega} = a(\omega) \hat{x} + \int_0^{\infty}
d\omega^{\prime}b_{\omega^{\prime}}(\omega) X_{\omega^{\prime}},
\end{equation}
where the coefficients, $a(\omega)$ and $b_{\omega^{\prime}}(\omega)$ are
assumed to be real. The coefficients are chosen to satisfy the eigenmode
equations
\begin{eqnarray}
& & \bar{\omega}_0^2 a(\omega) + \int_0^{\infty}
d\omega^{\prime}C_{\omega^{\prime}} b_{\omega^{\prime}}(\omega) = \omega^2
a(\omega),  \label{eq:Sch1} \\
& & C_{\omega^{\prime}} a(\omega) + {\omega^{\prime}}^2
b_{\omega^{\prime}}(\omega) = \omega^2 b_{\omega^{\prime}} (\omega),
\label{eq:Sch2}
\end{eqnarray}
where $\bar{\omega}_0^2 = \omega_0^2 + \int d\omega^{\prime}C_{{%
\omega^{\prime}}^2}/{\omega^{\prime}}^2$. Then, the Hamiltonian of the $\pi$%
-junction can be diagonalized as
\begin{equation}
H_{\pi} = \int_0^{\infty} d\omega \left( \frac{\bar{P}_{\omega}^2}{2} +
\frac12 \omega^2 \bar{X}_{\omega}^2 \right).  \label{eq:pi_junction}
\end{equation}
In order to solve the eigenmode equations, eqs.~(\ref{eq:Sch1}) and (\ref%
{eq:Sch2}), we may follow the calculation in Fano's paper~\cite{Fano61}. We
only give the result for $a(\omega)$ as
\begin{eqnarray}
|a(\omega)|^2 &=& \frac{C_{\omega}^2}{(\pi^2 C_{\omega}^4/4\omega^2) +
(\omega^2 - \bar{\omega}_0^2 - F(\omega))^2}  \label{eq:result} \\
F(\omega) &=& \mathrm{P} \int_0^{\infty} d\omega^{\prime} \frac{ C_{ {%
\omega^{\prime}}^2 }}{(\omega^2 - {\omega^{\prime}}^2)}.
\end{eqnarray}
The part of the energy renormalization is modified as
\begin{equation}
\bar{\omega}_0^2 + F(\omega) = \omega_0^2 + \mathrm{P} \int_0^{\infty}
d\omega^{\prime} \frac{\omega^2 C_{{\omega^{\prime}}^2}} {{\omega^{\prime}}%
^2(\omega^2 - {\omega^{\prime}}^2)}.
\end{equation}
Here, the second term in r.h.s. can be neglected, because it can be shown to
be $\mathrm{O}(\Delta/\omega_{\mathrm{c}})$.

Thus, we obtain the new expression for the $\pi $-junction as (\ref%
{eq:pi_junction}), while the coupling term (\ref{eq:coupling}) is rewritten
by the relation
\begin{equation}
x=\int_{0}^{\infty }d\omega ^{\prime }a(\omega ^{\prime })\bar{X}_{\omega
^{\prime }}.  \label{eq:x_trans}
\end{equation}%
As a result, we obtain the total Hamiltonian as
\begin{eqnarray}
H &=&\frac{\Delta }{2}\sigma _{x}+\frac{\varepsilon }{2}\sigma _{z}
\nonumber \\
&-&E_{J,\mathrm{eff}}(8E_{C,\pi })^{1/2}\sigma _{z}\int_{0}^{\infty }d\omega
a(\omega )\bar{X}_{\omega }  \nonumber \\
&+&\int_{0}^{\infty }d\omega \left( \frac{\bar{P}_{\omega }^{2}}{2}+\frac{1}{%
2}\omega ^{2}\bar{X}_{\omega }^{2}\right) .
\end{eqnarray}%
In this modified spin-Boson model, the effective spectral function is given
by
\begin{eqnarray}
&&J_{\mathrm{eff}}(\omega )=8E_{J,\mathrm{eff}}^{2}E_{C,\pi }\times \frac{%
\pi }{2}\frac{|a(\omega )|^{2}}{\omega }  \nonumber \\
&=&8E_{J,\mathrm{eff}}^{2}E_{C,\pi }\times \frac{(\pi C_{\omega
}^{2}/2\omega )}{(\pi C_{\omega }^{2}/2\omega )^{2}+(\omega ^{2}-\omega
_{0}^{2})^{2}}.  \label{eq:Jeff0}
\end{eqnarray}%
For the RSJ model, by substituting $J(\omega )/M=\pi C_{\omega
}^{2}/(2\omega )=\gamma \omega $, the effective spectral function is
obtained for $\omega \ll \omega _{\mathrm{c}}$ as
\begin{equation}
J_{\mathrm{eff}}(\omega )=8E_{J,\mathrm{eff}}^{2}E_{C,\pi }\times \left[
\frac{\gamma \omega }{\gamma ^{2}\omega ^{2}+(\omega ^{2}-\omega
_{0}^{2})^{2}}\right] .  \label{eq:Jeff_ex}
\end{equation}
Note that the form factor of the damped oscillator (the factor in the
bracket) appears in the effective spectral function.

By using the effective spectral function $J_{\mathrm{eff}}(\omega )$, we
estimate the dephasing time of the qubit at the optimal point ($\varepsilon =0$),
where long coherence time is realized by suppressing a linear coupling to the heat-bath.
The dephasing time is evaluated within the spin-Boson model in the form~\cite{Makhlin01,Leggett87}
\begin{equation}
\tau _{\varphi }^{-1}=\frac{1}{2}\tau _{\mathrm{relax}}^{-1} + \frac{1}{T_{2}^{\ast }}.
\label{eq:dephasing_form}
\end{equation}
The relaxation rate $\tau_{\mathrm{relax}}$ is calculated as
\begin{equation}
\tau _{\mathrm{relax}}^{-1}=2J_{\mathrm{eff}}(\Delta )\coth
\left( \frac{\Delta }{2k_{\rm B} T}\right).  \label{eq:tau_relax}
\end{equation}
On the other hand, $1/T_2^{\ast}$, which is a pure dephasing rate due to a quadratic coupling
to the heat-bath at the optimal point, is calculated from
the coupling strength $r = \lim_{\omega \rightarrow 0} J(\omega)/\omega$ as~\cite{Makhlin04}
\begin{eqnarray}
\frac{1}{T_{2}^{\ast }} &=& \frac{16\pi}{3} \left(\frac{r}{\Delta} \right)^2 (k_{\rm B} T)^3
\label{eq:T2form} \\
&=& \frac{\pi}{12}\left( \frac{E_{J,\mathrm{eff}}^{2} \gamma}{E_{J,\pi }^{2}E_{C,\pi }}\right) ^{2}
\frac{(k_{\mathrm{B}}T)^{3}}{\Delta ^{2}}  \label{eq:T_2}
\end{eqnarray}

We estimate the dephasing time in the present flux qubit by using the
parameters in Ref.~\onlinecite{Chiorescu03}. In the experiment, the
parameters are chosen as $E_J/k_{\mathrm{B}} = 12~\mathrm{K}$, $E_C/k_{%
\mathrm{B}} = 350~\mathrm{mK}$, $\Delta/k_{\mathrm{B}} = 160~\mathrm{mK}$, $%
T =25~\mathrm{mK}$, and $\alpha = 0.8$. There are several candidates of $\pi$%
-junctions for phase bias. We have estimated dephasing time for several $%
\pi$-junction systems, and found that only underdamped $\pi$-junctions may give
a sufficiently long dephasing time.

\begin{figure}[tb]
\begin{center}
\includegraphics[width=80mm]{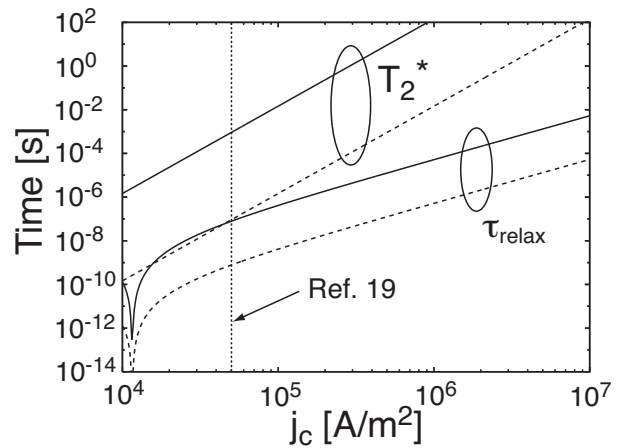}
\end{center}
\caption{The estimated relaxation time $\tau_{\rm relax}$ and pure dephasing time $T_2^{\ast}$.
The result for a $10~\mu\mathrm{m}\times10~\mu\mathrm{m}$ 
($1~\mu\mathrm{m}\times1~\mu\mathrm{m}$) $\pi$-junction is shown
by the solid (dashed) line.
}
\label{fig:result}
\end{figure}

Here, we discuss underdamped SIFS junctions by using the parameters
in Ref.~\onlinecite{Weides06}; We choose a capacitance and a subgap
resistance for unit area as $c = 0.08~\mathrm{F}/\mathrm{m}^2$ and
$r_{\mathrm{n}} = 3.0 \times 10^{-7}~\Omega
\mathrm{m}^2$~\cite{footnote1}, respectively. In
Ref.~\onlinecite{Weides06}, the measured critical current density is
$j_{\mathrm{c}} = 5.0 \times 10^4~\mathrm{A}/\mathrm{m}^2$. Here, we
take the critical current density as a parameter, and discuss its dependence
keeping $r_{\mathrm{n}}$ constant. In Fig.~\ref{fig:result}, we
show the relaxation time $\tau_{\mathrm{relax}}$ and the pure depasing time $T_2^{\ast}$
as a function of the critical current density
for a $1~\mu \mathrm{m} \times 1~\mu \mathrm{m}$ and $10~\mu
\mathrm{m} \times 10~\mu\mathrm{m}$ junction. In this estimate, the
relaxation process is always dominant ($\tau_{\mathrm{relax}} \ll T_2^{\ast}$), and therefore
the dephasing time $\tau_{\varphi}$ is determined by $\tau_{\mathrm{relax}}$.
We find that at the critical current density of
Ref.~\onlinecite{Weides06} the dephasing time is very short, while
long coherence time is obtained for junction with larger area and
higher critical current density. The relaxation time has a resonant structure
at a low critical current $j_{\mathrm{c}} =j_{\mathrm{c}}^{\ast}$, where
the resonant condition $\omega_0 = \Delta$ is satisfied. 
For $j_{\mathrm{c}} \gg j_{\mathrm{c}}^{\ast}$, 
the relaxation time and the pure dephasing time depend on the junction area $A$
and critical current  density $j_{\mathrm{c}}$ as
$\tau_{\mathrm{relax}} \propto j_{\mathrm{c}}^{2} A$ and $T_2^{\ast}
\propto j_{\mathrm{c}}^{4} A^{2}$, respectively. As
seen in Fig.~\ref{fig:result}, if we use underdamped $\pi$-junctions 
with large critical current ($j_{\mathrm{c}} \sim 10^7~\mathrm{A}
/\mathrm{m}^2$) and large junction area ($A \sim 10~\mu
\mathrm{m} \times 10~\mu \mathrm{m}$), coherence time becomes of
order of $1~\mathrm{ms}$, which is sufficiently long comparing to the
decoherence time limited by other sources. We note that 
when the relaxation process is dominant, the dephasing time is proportional to $E_{J, \pi}^2 R$;
for realization of long coherence time we need to increase both 
the Josephson energy $E_{J, \pi}$ and the subgap resistance $R$ of the
$\pi$-junction. 


Thus, for long-time coherent operations, one has to improve the quality factor by
changing experimental parameters of $\pi$-junctions. Especially important
parameter is the critical current density in the $\pi$-state of the junction. In the SIFS
junctions described in Ref.~\onlinecite{Weides06} the critical current density
$j_{\mathrm{c}}\simeq 5 \times 10^4 \mathrm{A}/\mathrm{m}^2$ in the $\pi$-state
was still rather low, three orders of
magnitude less than $j_{\mathrm{c}}\simeq 4 \times 10^7 \mathrm{A}/\mathrm{m}^2$ in
a SIS junction having the same tunnel barrier.
Possible reason for the strong suppression of the critical current
is the use of diluted alloy
Ni$_{x}$Cu$_{1-x}$ which has rather strong disorder leading to fast decay of
the supercurrent with increasing F-layer thickness. Since 0-$\pi$
transition occurs at certain critical thickness of the F-layer, the
supercurrent in the $\pi $-state is much smaller than in the 0-state. However,
smallness of  $j_{\mathrm{c}}$ is not an intrinsic property of SIFS
junctions. In a clean homogeneous ferromagnet the decay length may become much
longer than the 0-$\pi $ transition thickness. Recent experiments~\cite{Born06}
using Ni$_{3}$Al have demonstrated multiple 0-$\pi$
transitions with only modest decay of $|j_{\mathrm{c}}|$ as a function of
the thickness of Ni$_{3}$Al.
Therefore, choosing different materials for a ferromagnet layer
may finally lead to increasing $j_{\mathrm{c}}$ and thus to an increase of
the dephasing time of qubits with an SIFS junctions.

Finally, we discuss the advantage of the present phase bias. In
usual flux qubits, external magnetic flux is needed to produce phase
bias along the loop. 
In many experiments an external coil with a large current and weak
coupling to the qubits has been used. However, this prevents one from
using a superconducting shield which provides good shielding of qubits from
external flux noise. On the other hand, if one uses a local biasing with a
control line, noise in the current source degrades the coherence of the qubit.
From eq.~(\ref{eq:tau_relax}),
the relaxation time due to this noise can be evaluated as $\tau_{\mathrm{relax}}^{-1}
= (2\Delta/\hbar^2) (MI_{\mathrm{p}})^2/Z$, where $I_p = 2\pi E_{J, \mathrm{eff}}/\Phi_0$ 
is a circulating current, $M$ is a mutual inductance, and $Z$ is an
impedance of the current source~\cite{Makhlin01,vanderWal03}. 
For obtaining the dephasing time
longer than $1~\mathrm{ms}$, the maximum value of the mutual inductance $M$
is estimated to be $0.03~\mathrm{pH}$ for $Z=50~\Omega$. Then, the external
current needed for the phase bias becomes $40~\mathrm{mA}$, 
which is unrealistically high. Therefore, the use of
$\pi$-junctions may be an attractive option for individual phase
biasing on qubits.

Recently, another phase biasing scheme with a trapped flux in a
superconducting loop has been proposed and demonstrated~\cite{flux_trap}. 
A possible advantage of our scheme using a $\pi$-junction is that we do not
need to apply the large external field corresponding to a half flux in the loop ever,
either globally or locally. This makes implementation of a superconducting shield
simpler.

In summary, we proposed a simple phase bias by $\pi$-junctions for flux
qubits, and studied dissipation effects at the $\pi$-junction. In the framework of
the Caldeira-Leggett model, we derived the effective spectral function of
the spin-Boson model, and used it for estimate of the dephasing time of the
proposed qubit. We showed that for long coherent operation 
both the subgap resistance and critical current of the $\pi$-junction have to be
increased. We expect that further
improvement in quality of $\pi$-junctions enables us to use it for a $\pi$
phase shifter for flux qubits.

We also acknowledge to H. Hilgenkamp,
Ariando, K. Verwijs, A. Andreski, A. V. Ustinov, V. V. Ryazanov, A.
K. Feofanov, and S. Kawabata for helpful discussion. This work was
supported by the NanoNed Program under Project No.~TCS.7029, and
was partially supported by CREST, JST.


\begin{thebibliography}{99}
%

\bibitem{Tsuei94} C. C. Tsuei \textit{et al.}, Phys. Rev. Lett. \textbf{73}, 593 (1994)

\bibitem{Tsuei00} C. C. Tsuei and J. R. Kirtley, Rev. Mod. Phys. \textbf{72}, 969 (2000).

\bibitem{high_Tc_low_Tc} H. Hilgenkamp \textit{et al.}, Science \textbf{422}, 50 (2003);
H.-J. Smilde \textit{et al.}, Appl. Phys. Lett. \textbf{85}, 4091 (2004).

%
%

\bibitem{quasi_particle_injection}J. J. A. Baselmans \textit{et al.}, Nature \textbf{397}, 43 (1999);
J. J. A. Baselmans \textit{et al.}, Phys. Rev. Lett. \textbf{89}, 207002 (2002).

%
%

\bibitem{SFS} V. V. Ryazanov \textit{et al.}, Phys. Rev. Lett. \textbf{86}, 2427 (2001);
V. V. Ryazanov \textit{et al.}, Phys. Rev. B \textbf{65}, 020501(R) (2001);
T. Kontos \textit{et al.}, Phys. Rev. Lett. \textbf{89}, 137007 (2002);
H. Sellier \textit{et al.}, Phys. Rev. B \textbf{68}, 054531 (2003);
A. Bauer \textit{et al.}, Phys. Rev. Lett. \textbf{92}, 217001 (2004);
V. A. Oboznov \textit{et al.}, Phys. Rev. Lett. \textbf{96} 197003 (2006).

%
%

\bibitem{logic}
E. Terzioglu and M. R. Beasley, IEEE Trans. Appl. Supercond. \textbf{8}, 48 (1998);
A. V. Ustinov and V. K. Kaplunenko, J. Appl. Phys. \textbf{94}, 5405 (2003).

\bibitem{Ortlepp06} T. Ortlepp \textit{et al.}, Science \textbf{319}, 1495 (2006).

%
%

\bibitem{quiet_qubit}L. B. Ioffe \textit{et al.}, Nature \textbf{398}, 679 (1999);
G. Blatter \textit{et al.}, Phys. Rev. B, \textbf{63}, 174511 (2001).

%
%

\bibitem{charge_qubit}
Y. Nakamura \textit{et al.}, Nature \textbf{398}, 786 (1999);
T. Yamamoto \textit{et al.}, Nature \textbf{425}, 941 (2003);
A. Wallraff \textit{et al.}, Phys. Rev. Lett. \textbf{95}, 060501 (2005).

%
%

\bibitem{quantronium_qubit}
D. Vion \textit{et al.}, Science \textbf{296}, 886 (2002);
I. Siddiqi \textit{et al.}, Phys. Rev. B, \textbf{73}, 054510 (2006).

%
%

\bibitem{flux_qubit}
J. E. Mooij~\textit{et al.}, Science \textbf{285}, 1036 (1999);
C. H. van der Wal~\textit{et al.}, Science \textbf{290}, 773 (2000);
I. Chiorescu~\textit{et al.}, Nature \textbf{431}, 159 (2004);
A. Lupascu et al. Nature Phys. \textbf{3}, 119 (2007)

\bibitem{Chiorescu03}
I. Chiorescu~\textit{et al.}, Science \textbf{299}, 1869 (2003);

%
%

\bibitem{phase_qubit}
J. M. Martinis \textit{et al.}, Phys. Rev. Lett. \textbf{89}, 117901 (2002);
R. McDermott \textit{et al.}, Science \textbf{397}, 1299 (2005).

%
%

\bibitem{Fano61} U. Fano, Phys. Rev. \textbf{124}, 1866 (1961).

\bibitem{Costa00} M. Rosenau da Costa~\textit{et al.}, Phys. Rev. A \textbf{61} 022107 (2000).

%
%

\bibitem{Makhlin01} Y. Makhlin~\textit{et al.}, Rev. Mod. Phys. \textbf{73}, 357 (2001).

\bibitem{Leggett87} A. J. Leggett~\textit{et al.}, Rev. Mod. Phys. \textbf{59}, 1 (1987).

\bibitem{Makhlin04} Y. Makhlin and A. Shnirman, Phys. Rev. Lett. \textbf{92}%
, 178301 (2004). 

\bibitem{Weides06} M. Weides \textit{et al.}, Appl. Phys. Lett. \textbf{89}, 122511 (2006).

\bibitem{footnote1}We have estimated the subgap resistance from the McCumber
parameter 
$\beta = 2\pi I_{\mathrm{c}}R^2 C/\Phi_0$ measured in Ref.~\onlinecite{Weides06}. We have chosen
$\beta = 10^6$ by extrapolating the data of Ref.~\onlinecite{Weides06} to the
low-temperature region.

\bibitem{Born06}F. Born~\textit{et. al.}, Phys. Rev. B \textbf{74}, 140501(R) (2006).

\bibitem{vanderWal03}C. H. van der Wal \textit{et al.}, Eur. Phys. J. B \textbf{31}, 111 (2003).


\bibitem{flux_trap} J. B. Majer~\textit{et al.}, Appl. Phys. Lett. \textbf{80}, 3638 (2002);
J. H. Plantenberg \textit{et al.} Nature \textbf{447}, 836 (2007).
\end{thebibliography}
\end{document}